\documentclass[aps,prl,onecolumn,floats]{revtex4}
\usepackage{latexsym}
\usepackage{amsmath}
\usepackage{amssymb}
\usepackage{bm}
\usepackage{wasysym}
\usepackage[dvips]{color}

\usepackage{graphicx}
\usepackage{graphicx}
\usepackage{subfigure}
\usepackage{epsfig}

\newcommand{\hide}[1]{}
\newcommand{\veps}{\varepsilon}
\newcommand{\la}{\langle}
\newcommand{\ra}{\rangle}

\begin{document}

\title{Feshbach Resonance without a Closed Channel Bound-State} 

\author{Y. Avishai} 
\affiliation{Department of Physics and the Ilse Katz Center
for Nano-Science, Ben-Gurion University, Beer-Sheva 84105, Israel, and\\
Department of Physics, Hong Kong University of Science and
Technology, Kowloon, Hong Kong} 

\author{Y. B. Band}
\affiliation{Department of Chemistry, Department of Physics and
Department of Electro-Optics, and the Ilse Katz Center for
Nano-Science, Ben-Gurion University, Beer-Sheva 84105, Israel}

\author{M. Trippenbach}
\affiliation{Institute of Theoretical Physics, University of
Warsaw, ul.  Ho\.{z}a 69, PL--00--681 Warszawa, Poland}

% \date{\today}

\begin{abstract}
The physics of Feshbach resonance is analyzed using an analytic
expression for the $s$-wave scattering phase-shift and the scattering
length $a$ which we derive within a two-channel tight-binding model.
Employing a unified treatment of bound states and resonances in terms
of the Jost function, it is shown that for strong inter-channel
coupling, Feshbach resonance can occur even when the closed channel
does not have a bound state.  This may extend the range of ultra-cold
atomic systems that can be manipulated by Feshbach resonance.  The
dependence of the sign of $a$ on the coupling strength in the unitary
limit is elucidated.  As a by-product, analytic expressions are
derived for the background scattering length, the external magnetic
field at which resonance occurs, and the energy shift $\veps-\veps_B$,
where $\veps$ is the scattering energy and $\veps_B$ is the
bound state energy in the closed channel (when there is one).
\end{abstract}


\maketitle

{\it Introduction.}--- Feshbach resonance (FR) enables manipulation of
the interactions between ultra-cold atoms, e.g., it allows a repulsive
gas to be transformed into an attractive one and vice-versa
\cite{Feshbach, Kerman, Julienne1, CCT, Chin_10, Blatt_11, Jones_06},
as in a BEC-BCS crossover \cite{Regal_04}.  The paradigm of FR, as in
the low energy collision of two ultra-cold atoms, involves the
coupling of an open channel, $o$, and a bound state at energy
$\veps_B$ in (another) closed channel, $c$, giving rise to resonant
variations of the $s$-wave scattering length $a$ \cite{CCT}.  This
resonance occurs when $\veps_B \approx \veps$ where $\veps$ is the
scattering energy, namely, when the energy of the closed channel bound
state is close to the threshold of the open channel.  This condition
can be experimentally implemented by varying an external parameter,
such as a static magnetic field $B$, so that the bound state energy is
swept through resonance.  As we show below, this paradigm can be
extended to the case where the closed channel does not have a bound
state.

FR is usually formulated as a two-channel scattering problem, which
establishes the relation between the bare parameters of the scattering
problem (electronic potentials, coupling strength, scattering energy
and external fields) and the physically relevant observables (e.g.,
the scattering length versus external field, $a(B)$, the bound states
of the coupled system, etc.).  In this context, it is useful to
consider a two-channel model that allows the derivation of analytical
expressions of these observables in order to elucidate their relation
to the bare parameters.  Such a model is not intended to analyze a
specific system in detail in which the intra-channel potentials and
the inter-channel coupling have a definite form appropriate for the
system under study.  Rather, it should be sufficiently general and
simple and, at the same time, encode the underlying physics.  The
basic ingredients of a two channel $s$-wave scattering problem
designed to analyze a FR include: (1) The intra-channel potentials
$v_o(r)$ of the open channel and $v_c(r)$ of the closed channel (here
$r$ is the distance between the two atoms).  (2) The inter-channel
coupling potential $w(r)$ (for simplicity we take a constant coupling
strength $\tau$).  (3) An external tunable parameter controlling the
energy difference $v \equiv v_c(r)-v_o(r)$ as $r \to \infty$.
Experimentally, $v$ is often tuned by varying an external magnetic
field, $v = \alpha B$, where the constant $\alpha$ depends on the
specific system.  Thus, knowing the value $v=v_0$ for which the system
has a FR is equivalent to finding the magnetic field $B_0$ at which
the scattering length is infinite.

Our main objective here is to show that FR can occur even when the
closed channel {\it does not have a bound state}, or even when the
atom-atom potential in this channel is repulsive.  The motivation for
addressing this question is evident: this will demonstrate that
systems for which a transition from an attractive gas of atoms to a
repulsive gas are feasible even for systems for which $v_c(r)$ does
not support a bound state.  Using a simple model, we show that this is
indeed the case; it is possible to obtain a FR and a bound state of
the coupled-channel system for large enough coupling $\tau$ of the
closed and open channels, even when there is no bound state in the
closed channel.  As a by product, analytic expressions are derived for
the basic physical observables related to FR in terms of the
parameters of the scattering problem and a unified treatment of bound
states and resonances is carried out in terms of the Jost function.

%--------------------continuum model-----------------------
{\it The two-channel scattering problem.}--- The $s$-wave two-channel
scattering problem between two atoms a distance $r$ can be mapped onto
a single-particle scattering problem in the center of mass coordinate
system governed by the Schr\"odinger equation (in abstract form), 
\begin{equation} \label{Eq:SE}
    \binom{H_o \ w^\dagger}{\ w \ \ H_c} \binom {u_o}{u_c} = \veps \binom
    {u_o}{u_c}=k^2 \binom {u_o}{u_c} ~.
\end{equation}
 $H_o = -\tfrac{d^2}{dr^2} + v_o(r)$ and $H_c =
-\tfrac{d^2}{dr^2} + v_c(r)$ are the Hamiltonians (in $r$ space) for the open and
closed channels composed of the kinetic energy operator and
intra-channel potentials $v_o$ and $v_c$, and $\veps=\veps_k =k^2$.  The
open and closed channels are coupled
 by the potential
$w(r)$.  The boundary conditions satisfied by the closed and open
components of the exact wave function are $u_o(0) = u_c(0)=0$,
$u_c(r)_{r \to \infty}= A e^{-\kappa r}$, and $u_o(r)_{r \to \infty} =
B \sin [kr$+$\delta(k)]$.  Here $\kappa(\veps_k)>0$, $A$ and $B$ are
energy dependent constants and $\delta(k)$ is the scattering phase
shift.  The $s$-wave scattering length is given by
\begin{equation} \label{a}
    a = -\lim_{k \to 0} \frac{\tan \delta(k)}{k},
\end{equation}
In the standard picture of FR, the closed channel, when uncoupled from
the open channel, is assumed to have a bound state $|B \ra$ at energy
$\veps_B<v_c(\infty)$ and continuum states, $\{|p \ra \}$ at energies
$\{ \epsilon_p > v_c(\infty) \}$.  The scattering states of $H_o$ are
defined as $H_o|k \ra=\veps|k \ra=k^2|k \ra$ (assuming $v_o(\infty)=0$
for simplicity).  Eliminating $|u_c \ra$ from the set of coupled
equations (\ref{Eq:SE}) results in a single equation for $|u_o \ra$,
$[H_o + v_{\mathrm{eff}}(\veps)]|u_o\ra = \veps |u_o\ra,$ with an
effective potential, $v_{\mathrm{eff}}(\veps) = wG_c(\veps)w^\dagger$, where
$G_c(\veps) = (\veps - H_c)^{-1}$.  The $T$ matrix associated with
$v_{\mathrm{eff}}(\veps)$ is formally given by $T(\veps)=[1-G_o(\veps)
v_{\mathrm{eff}}(\veps)]^{-1}$, and, by definition, $a=-C \lim_{k \to
0} \la k|T(\veps)|k \ra$, where $C>0$ is a kinematic constant.

For example, a particularly simple model takes $v_o(r)$ and $v_c(r)$
to be spherical square wells of range $R$, while $w(r)$ couples the
two channel only at $R$.  Explicitly,
\begin{eqnarray} \label{Eq:supp}
    && v_c(r)=v \Theta(r-R)+(v-\Delta) \Theta(R-r) \nonumber \\
    &&v_o(r)=-\Lambda \Theta(R-r), \ \ w(r) = \tau \, \delta(r-R),
\end{eqnarray}
where $\tau$ is the coupling strength.  Despite being a simple model,
an exact solution of the scattering problem requires solving a set of
coupled transcendental equations.  Its numerical solution
\cite{supp_mat} confirms the analytical results obtained within the
tight-binding (TB) model to which we now turn.

%-------------Tight-Binding model-----------------
{\it Tight-Binding Model.}--- Starting from the continuous model, we
discretize the radial coordinate, $r \to n$, where $n>0$ is an
integer, and replace $- \frac{d^2}{dr^2}$ by a second-order difference
operator \cite{Molmer}.  In second quantization it translates as a
hopping term, $- (\sum_{n \ge 1}a_n^\dagger a_{n+1}+h.c.)$, where
$a_n$ and $a_n^\dagger$ are the annihilation and creation operators of
the scattered particle on the positive integer grid-sites $n>0$.  The
potentials are,
\begin{align}
v_c(n)  = (v - \Delta) \delta_{n,1} + v \Theta(n - 1), &&  \nonumber \\
v_o(n) = -\Lambda \delta_{n,1}, \quad w(n) = \tau \delta_{n,1}.  &&
\end{align}
After treating the closed and open channels separately, we will solve
the coupled-channel scattering problem.

%---------------------The closed channel----------------
The Hamiltonian of the closed channel is,
\begin{equation} \label{H1}
    H_c = (v-\Delta) a_1^\dagger a_1 + \sum_{n > 1}[v \, a^\dagger_n
    a_n -(a_n^\dagger a_{n-1}+h.c.)] .
\end{equation}
If the well-depth $\Delta > 0$, the potential is attractive at site
$n=1$.  The potential height $v$ for $n>1$ is experimentally tunable
(e.g., via magnetic field).  Let $f_n$ be the amplitude of the wave
function at site $n$.  For a bound state with binding energy
$\veps_B$, $f_n = A_c \, e^{-\kappa (n-2)} \ (n>1)$ with $\kappa > 0$
and $A_c$ is a constant.  Therefore, for $n>1$ (i.e., outside the
range of the attractive potential),
%\begin{equation} \label{3}
$f_2=A, \ \ f_3 = A \, e^{- \kappa}$ and $\veps_B =v -2 \cosh\kappa)$.
%\end{equation}
Simple manipulations yield,
\begin{equation} \label{4}
    \kappa=\log \Delta >0 ~ \Rightarrow \Delta \ge 1, \quad \veps_B =
    v-\left [ \Delta+\frac{1}{\Delta} \right ].
\end{equation}
Thus, there is a threshold potential depth $\Delta >1$ for having an
$s$-wave bound state.  A similar scenario occurs also in a 3D
continuous geometry, in contradistinction to symmetric 1D or 2D
potentials, where {\it any} attractive potential of whatever strength
supports a bound state.  In the model treated here, at most one bound
state can occur \cite{Wasak_13}.  An artifact of the TB model is that
for a repulsive potential with $\Delta<-1$, the closed channel does
have a bound state above the upper band edge \cite{Manuel}.  To avoid
this, we will restrict the potential depth to $\Delta > - 1$.  To
summarize, for $\Delta > 1$ the closed channel has a bound state while
for $1 > \Delta > 0$, $v_c$ is attractive but there is no bound state.
Moreover, for $0 > \Delta >-1$, $v_c$ is repulsive and there is no
bound state.

%------------Open channel----------------
For the open channel we use $b^\dagger_n$ and $b_n$ as creation and
annihilation operators.  The Hamiltonian is,
\begin{equation} \label{HO}
   H_o = -\Lambda b_1^\dagger b_1-\sum_{n \ge 1} (b^\dagger_{n+1}b_n +
   h.c.).
\end{equation} 
For $\Lambda > 1$ the open channel has a bound state, for $1 > \Lambda
>0$, $v_o$ is attractive, but there is no bound state, whereas for $0
> \Lambda > -1$, $v_o$ is repulsive and there is no bound state.  The
wave function on site $n \ge 1$ is $g_n = A_o \sin [k
(n-1)+\delta(k)]$, where $A_o$ is a constant.  The continuous spectrum
is a band of energies,
\begin{equation} \label{5}
    \varepsilon_k = -2 \cos k~, \ \Rightarrow \ -2 \le \varepsilon_k
    \le 2,
\end{equation}
so that the lowest threshold for propagation is $\varepsilon_{k=0}=-2$.

%--------------------FR--------------------------------------
Now, consider the coupled-channel system.  The Hamiltonian is,
\begin{equation} \label{H12}
    H = H_c+H_o+\tau(a_1^\dagger b_1+h.c.)~.
\end{equation}
In a scattering scenario, the effective particle approaches the
``origin'', $n=0$, in the open channel from right to left at a given
energy, $\varepsilon_k = -2 \cos k$, and is reflected back (rightward)
into the open channel.  The reflection amplitude, equivalently the $S$
matrix, is $S=e^{2 i \delta(k)}$, where $\delta(k)$ is the $s$-wave
phase shift from which $a$ is computed as in Eq.~(\ref{a}).  $f_n$ and
$g_n$ are the amplitudes of the wave function on site $n$ for the
closed and open channels respectively [analogous of $u_c(r)$ and
$u_o(r)$ in the continuous model].  The ``asymptotic" forms of $f_n$
and $g_n$ are,
\begin{equation} \label{6} 
    f_n=A_ce^{-\kappa n}, \ \ g_n=\sin [k (n-1)+\delta(k)], \ \ (n>1),
\end{equation}
where $\kappa$ is related to $\varepsilon_k$ as, $ 2\cosh
\kappa=v-\varepsilon_k \ge 2$.  Thus, the ``non-leakage" condition, $v
- \varepsilon_k > 2$, guarantees that propagation in the closed
channel is evanescent.  Unlike Eq.~(\ref{4}), here $\kappa$ is
independent of the depth $\Delta$ of $v_c$.

%------------------result 1------------------
{\it Solution.}---
Solving the TB equations we obtain a relatively simple expression for
$\tan \delta(k)$, independent of sgn$(\tau)$.  Writing $\tan
\delta(k)=N/D$ and $q(v,k) = \sqrt{(v-\veps_k)^2-4}$, we find
\begin{eqnarray} \label{DtvDelta}
  && N(k,\tau,v,\Delta,\Lambda) = \{ 2
  \tau^2+\Lambda[v-\veps_k+q(v,k)-2 \Delta] \} \sin k, \nonumber \\
  && D(k,\tau,v,\Delta,\Lambda)= q(v,k)+v - 2\Delta+(\tau^2-1)
  \varepsilon_k \nonumber \\
  && \quad + \Lambda [(2 \Delta - v) \cos k-q(v,k) \cos k -\cos 2k -1] .
\end{eqnarray}
Because $N(k,...)=-N(-k,...)$ and $D(k,...)  = D(-k,...)$, $\delta(-k)
= -\delta(k)+n \pi$ ($n=$ integer).  Extracting $\delta(k)$ from $\tan
\delta(k)$ requires a reference; in the continuum version,
$\delta(\infty) = 0$, but in TB, ``$\infty$'' refers to $k=\pi$.

%----------tuning v---------------------
Next we find at what potential $v=v_0(\tau,\Delta, \Lambda)$ we arrive
at a FR as $k \to 0$.  This is equivalent to finding the value of the
magnetic field $B_0$ for which there is a FR and $|a| \to \infty$ (an
experimentally relevant challenge).  Because $N(0,...)=0$, a necessary
condition on $v_0$ for achieving $|a|=-\lim_{k \to 0} [|\tan
\delta(k)|/k] = \infty$ is $D(0,\tau,v_0,\Delta,\Lambda) = 0$.
From Eq.~(\ref{DtvDelta}) we easily obtain,
\begin{equation} \label{v0}
    v_0(\tau, \Delta, \Lambda) =
    \frac{[\Lambda-1+\tau^2+\Delta(1-\Lambda)]^2}
    {[\Delta(1-\Lambda)+\tau^2](1-\Lambda)} .
\end{equation} 
Equation (\ref{DtvDelta}) is also a sufficient condition for
$|a|=\infty$ because $D(k,\tau,v_0,\Delta,\Lambda)$ vanishes as $k^2$
when $k \to 0$.  Therefore, when $v=v_0$, the denominator $D$ in
Eq.~(\ref{DtvDelta}) vanishes {\em faster than the numerator $N
\propto \sin k$} (see Sec.~III of the supplementary material), and
thus, $|a| \to \infty$.  Hence, for a given $\tau, \Delta, \Lambda$,
one can tune $v \to v_0(\tau,\Delta, \Lambda)$, Eq.~(\ref{v0}), in
order to achieve a FR [equivalently, $|\tan \delta(0)| = \infty$].

The discussion following Eq.~(\ref{H12}) dictates that $v_0$ must be
positive in order to guarantee the non-leakage condition of the closed
channel as discussed after Eq.~(\ref{6}).  Inspecting the expression
(\ref{v0}) for $v_0$, we see that it is reasonable to constrain
$\Lambda < 1$.  Under this condition, the open channel potential is
attractive (equivalently, $\Lambda >0$) but it does not support a
bound state ($\Lambda < 1$).

In order to substantiate our main result we need to understand the
relationship between the occurrence of bound states in a coupled-channel
system and FRs.  A uniform treatment of resonances and bound states of
the coupled-channel system is achievable in terms of the Jost
function.  For $v \ne v_0(\tau, \Delta, \Lambda)$ [Eq.~(\ref{v0})],
resonances and/or bound states of the coupled system exist for
$\varepsilon \ne -2$.  To explore this regime we use the Jost
function, defined (for fixed $\tau, \Delta, \Lambda$) as,
\begin{equation} \label{Jost}
    f(k,v) \equiv D(k,\tau,v, \Delta, \Lambda) - i N(k,\tau,v, \Delta,
    \Lambda).
\end{equation}
The $S$ matrix is given in terms of the Jost function as,
$S = e^{2i \delta} = (1+i \tan \delta)/(1-i \tan \delta) =
f(-k,v)/f(k,v)$.
%recalling that $D(k,...)  = D(-k,...)$ and $N(-k,...)  = -N(k,...)$.  
The Jost function in ordinary potential scattering is discussed in
textbooks on scattering theory, e.g. \cite{Taylor}, but here it is
formulated and exactly calculated for a `non-ordinary' scattering
problem with an effective energy-dependent potential.  Considered as a
function of the {\em complex variable} $z=k+iq$ ($0 \le k \le \pi, \
-\infty < q < \infty$), $f(z,v)$, is well defined on the imaginary $z$
axis, $z = iq$, where it is real.  Solving $f(iq,v)=0$ gives the
position $q(v)$ that is a pole of the $S$ matrix on the imaginary axis
in the $z$ plane.  An $s$-wave bound state appears as an isolated zero
of $f(i q, v<v_0)$ with $q>0$, whereas an $s$-wave resonance appears
as an isolated zero of $f(i q, v> v_0)$ with $q<0$.  In both cases,
the energy equals $-2 \cosh q < \veps_0=-2$ (namely, below the
continuum threshold), but, strictly speaking, the resonance energy is
located on the second Riemann sheet in the complex {\it energy} plane.
Finally, a FR is a zero of $f(i q,v_0)$ occurring at $q \to 0$.  Thus,
a small upward shift of $v_c$ turns a zero-energy bound state at
$q=0^+$ into a zero-energy (Feshbach) resonance at $q=0^-$.

%\begin{figure}[!ht]
%\centering
%\includegraphics[scale=0.5,angle=0]{Fig2.eps}
%\caption{Zeroes of Jost function $f(iq,v)$.  From left: bound state at
%$q>0, v<v_0$, FR at $q=0, v=v_0$ and virtual state at $q<0, v>v_0$.
%See Ref.~\cite{Taylor}.}
%\label{FigJost}
%\end{figure}

\begin{figure}[!ht]
\centering
\includegraphics[width=0.40 \textwidth,angle=0]{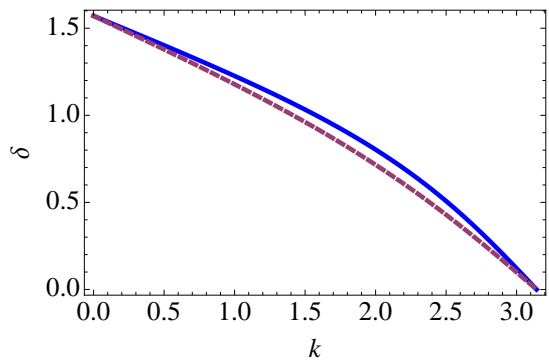}
\caption{Phase shift $\delta(k)$ versus $k$, Eq.~(\ref{DtvDelta}).
Solid curve: $\tau=2, \ \Delta=1.1 >1, \Lambda=0.2$ (the closed
channel has a bound state).  Dashed curve: $\tau=2.227, \Delta=-0.1
<0, \Lambda=0.2$ (the atom-atom potential in the closed channel is
repulsive).  In both cases $v_0(\tau, \Delta, \Lambda)$ defined in
Eq.~(\ref{v0}) is equal to 4.264 and $\delta(0)=\pi/2$, implying a
FR. The curves are virtually identical for small $k$ because $v_0$ is
determined by the phase shift near $k=0$.}
\label{Fig1}
\end{figure}

We are now in a position to derive our main result.
Equation~(\ref{v0}) shows that for fixed $0< \Lambda <1$, it is
possible to modify $\Delta \to \Delta'$ and $\tau \to \tau'$ in such a
way that $\tau'^2=\tau^2+(\Delta-\Delta')(1-\Lambda)$, without
affecting $v_0$.  We employ this property for the case $\Delta >1$
(the closed channel has a bound state) and $\Delta' <1$ (no
bound state in the closed channel), or even $-1 < \Delta' <0$.  The
equality $v_0(\tau,\Delta,\Lambda) = v_0(\tau',\Delta',\Lambda)$
guarantees that in both cases FR exists as is evident from
Fig.~\ref{Fig1} and as is explained in the caption.  However, only the
case $\Delta >1$ is commensurate with the paradigm of the FR spelled
at the introduction, according to which a bound state in the closed
channel is responsible for FR.

This somewhat unexpected result is not an artifact of the TB model.
In the supplementary material we present a formal proof for a
continuous model and substantiate it numerically.  To get a simpler
(albeit intuitive) physical picture, consider the equation for the
open channel $[H_o+v_{\mathrm{eff}} (\veps)] \psi_o = \veps \psi_o$
where $v_{\mathrm{eff}}(x) = w (x-H_c)^{-1} w^\dagger$.  If $H_c$ has
only a continuous spectrum starting at $\epsilon_0>0$ (i.e., $H_c$
{\it does not have a bound state}), then for every $x<0$ we have
$v_{\mathrm{eff}}(x) = -w[\sqrt{H_c-x} \sqrt{H_c-x}]^{-1} w^\dagger$
that is a negative definite hermitian operator.  Thus,
$v_{\mathrm{eff}}(x)$ is a non-local {\it attractive} potential.  By
properly tuning the coupling strength $\tau^2$ it is possible to get a
zero-energy bound state for $x=\veps=0^{-}$.
%consider coupled-channel scattering in the continuous case given in
%Eq.~(\ref{Eq:SE}), and show that if both $H_o$ and $H_c$ have only
%continuous spectra
%%(with respective 
%%states $|k \ra$ and $|p \ra$ and respective energies $\veps_k \ge 0$ and 
%%$\epsilon_p \ge v >0$) 
%then, for large enough coupling $\tau$, the equation for the open
%channel $[H_o+v_{\mathrm{eff}}]\psi_o = \veps \psi_o$ has a
%bound state solution at energy $\veps(\tau)<0$.  Moreover, $\tau$ can
%be tuned to achieve a zero-energy bund-state at energy $\veps = 0^-$.
As indicated in the discussion of the Jost function above, a small
upward shift of the closed channel potential $v_c$ moves a zero-energy
bound state into a zero-energy resonance, i.e., a FR. Summarizing, the
physical picture of this new type of FR is as follows: A strong
coupling leads to a zero-energy bound state in the equation of the
open channel alone (that includes an attractive potential
$v_{\mathrm{eff}}$).  Then, a slight upward shift of the closed
channel potential turns this zero-energy bound state into a FR.

Right at FR, $|a|=\infty$.  Properties of a unitary gas, where $|a|
\approx \infty$, are of great interest in cold atom physics
\cite{Castin}.  Direct analysis of Eq.~(\ref{DtvDelta}) (see Sec.~III
of the supplementary material) shows that the unitary gas is
attractive or repulsive depending upon the coupling strength $\tau$.
Specifically, there exists a threshold $\tau_0$ such that at the FR,
$a = -\infty \ (+\infty)$ for $\tau>\tau_0$ ($\tau < \tau_0$).

%Using formal scattering theory for the continuous
%model (\ref{Eq:SE}) we can show that a Feshbach resonance can occur
%even when both $H_c$ and $H_o$ have only continuous spectrum.
%The $T$ matrix derived from the effective potential $v_{\mathrm
%{eff}}$, (recall that $a \propto \lim_{k \to 0} \la k|T|k \ra$)
%includes a factor\\
%%\begin{equation}
%$ ~~~~~~~~ (1 -Q)^{-1} \equiv
%     [1 - (\veps-H_o)^{-1} w(\veps - H_c)^{-1}
%    w^\dagger]^{-1} .$\\
%    % \nonumber
%%\end{equation}
%In the supplementary material~\cite{supp_mat} we show that for $\veps
%\to 0$, $Q \equiv G_o(\veps)v_{\mathrm{eff}}(\veps)$ acting in ${\cal
%H}_o$ (the open channel Hilbert space spanned by the eigenstates of
%$H_o$) is positive definite.  Hence, by proper tuning of $w$, the $T$
%matrix can be manipulated to be singular at $\veps \to 0$, and since
%$a \propto \la k|T|k \ra$, we can obtain a zero-energy FR,
%$|a|=\infty$.

Finally, we use our analytic results within the TB model to derive
explicit expressions for several important quantities related to FR.
The simple expressions for these quantities teach us directly how
these physical observables depend on the parameters $v, \Delta,
\Lambda$ and $\tau$.  \\
(1) The functional dependence of $a$ on $v$ (that is proportional to
the applied magnetic field $B$) is of utmost importance.  Using
expression (\ref{DtvDelta}) and the definition (\ref{a}) we
immediately obtain,
\begin{equation} \label{av}
    a(v) = \frac{2 \tau^2+\Lambda [2(1-\Delta)+v+q(v,0)]}{2
    [1-\Delta(1-\Lambda)- (\tau^2+\Lambda)] +[v+q(v,0)](1-\Lambda) } .
\end{equation} 
For $v \approx v_0$, $a(v)$ diverges as $a(v) \propto 1/(v-v_0)$ where
the proportionality constant is easily calculated.  \\
(2) Another important quantity is the magnetic field $B_1$ at which
the scattering length vanishes, and $a$ changes sign without being
singular (recall that $v=\alpha B$).  For an atomic gas, whose
interaction is given to lowest order by a pseudo-potential, this means
a change from a repulsive gas to an attractive one (or vice versa).
$v_1$, the solution of $a(v_1)=0$ is given by, $v_1 =
-\frac{[\Lambda(1-\Delta)+\tau^2)]^2}{\Lambda(\tau^2-\Delta
\Lambda)}$.  Due to the non-leakage condition we must have $v_1>0$.
Hence, $a(v)$ vanishes only when $\Delta \Lambda > \tau^2$.  \\
(3) It is sometimes useful to partition the $s$-wave scattering length
into two terms, $a = a_{\mathrm{bg}} + a_{\mathrm{res}}$ where
$a_{\mathrm{bg}}$ is the contribution from the open channel alone and
$a_{\mathrm{res}}$ is the contribution due to coupling between the
closed and open channels.  By definition, $a_{\mathrm{bg}}=-\lim_{k
\to 0} [\tan \gamma(k)/k]$ where $\gamma(k)$ is the phase shift for
scattering from the open-channel potential $v_o(n)=-\Lambda
\delta_{n,1}$.  The result is,
\begin{equation} \label{abg}
    \tan \gamma(k) = \frac{\Lambda \sin k}{1-\Lambda \cos k} \
    \Rightarrow \ a_{\mathrm{bg}}=-\frac{\Lambda}{1-\Lambda}~.
\end{equation} 
(4) Of special interest is the energy shift $\Delta \veps
=\veps_B-\veps_0$ between the bound state energy of the closed channel
(when uncouple) and the scattering energy.  $\veps_0=-2$ at threshold.
A FR occurs when $v \to v_0$ as defined in Eq.~(\ref{v0}).  For
$v=v_0$ the closed channel, when uncoupled, has a bound state at
energy $\veps_B=v_0-(\Delta+1/\Delta)$ [for $\Delta>1$, see
Eq.~(\ref{4})].  Using this result implies,
\begin{equation} \label{Bplus2}
    \Delta \veps =
    \veps_B+2=\frac{\tau^2[(1-\Lambda)(\Delta^2-1)+\Delta
    \tau^2]}{(1-\Lambda)[\Delta(1-\Lambda)+\tau^2]}~.
\end{equation}
For $\Delta>1$ and $\Lambda <1$ we have $\Delta \veps>0$.  \\

%-----------------------Acknowledgment----------------
{\it Acknowledgement.} This work was supported in part by grants from
the Israel Science Foundation (Grant Nos.~400/2012 and 295/2011), the
James Franck German-Israel Binational Program, and a National Science
Center grant (M.T.).  We gratefully acknowledge discussions with H. A.
Weidenm\"uller and M. Valiente.

%------------------------Bibliography----------------- 

%\end{document}
\newpage
%\begin{multicols}{1}
%\documentclass[aps,prl,floats]{revtex4}

%% fonts
%\usepackage{latexsym}
%\usepackage{amsmath}
%\usepackage{amssymb}
%\usepackage{bm}
%\usepackage{wasysym}
%\usepackage[dvips]{color}

%\usepackage{graphicx}
%\usepackage{graphicx}
%\usepackage{subfigure}
%\usepackage{epsfig}

%\newcommand{\hide}[1]{}
%\newcommand{\veps}{\varepsilon}
%\newcommand{\la}{\langle}
%\newcommand{\ra}{\rangle}

%\begin{document}

%% \begin{widetext}
\begin{center}

{\textsf{\textbf{\large{Supplementary Material for ``Feshbach
Resonance without a Closed Channel Bound-State''}}}}

\end{center}
\bigskip

Here we reinforce the main of the Letter, that a zero energy Feshbach
resonance (FR) and a bound-state of the coupled (closed and open
channel) system occur also when the closed channel does not have a
bound-state.  This material contains two parts that complement each
other.  First we consider the model described by Eq.~(3) of the Letter
(albeit with $\Lambda=0$) and solve it exactly, arriving at the main
result as claimed above.  Then we treat a general model of coupled
channel $s$-wave scattering and show that if the spectra of the open
and closed channels have only continuous parts, the on-shell matrix
element of the scattering matrix is singular at zero energy, implying
the existence of a FR even when there is no bound-state in the closed
channel.  These two complementary analyses provide another
substantiation of this unexpected result, and show that it is not an
artifact of the tight-binding model.

In the last section we elaborate upon the relation between the sign of
the $s$-wave scattering length at resonance and the strength of the
coupling between the open and closed channel, as discussed in the main
text.

\section{I. Solution of the model defined by Eq. (3) with $\Lambda=0$}

We consider the exact wave function whose open and closed channel
components, $g(r)$ and $f(r)$, vanish at the origin, $r=0$.  The
potential of the open channel is zero, and the closed channel
potential is given by
\begin{equation} \label{SC1}
v_c(r)=\begin{cases} v >0 \ \ (r>R), \\ u=v-\Delta \ \ (r<R). \ \ \end{cases}
\end{equation}
where we use units so that $\hbar = 2 m = 1$.  Here $\Delta$ is the
depth of the closed channel potential, that is fixed by the atom-atom
potential.  The asymptotic closed channel potential $v$ is
experimentally tunable.  The two channels are coupled by a potential
$w(r)=\tau \delta(r-R)$ of strength $\tau$ at point $r=R$.

Within the $s$-wave scattering problem, a particle in the
open channel approaches the origin $r=0$ at scattering energy
$\veps=k^2$ and scattered back after gaining a phase-shift
$\delta(k)$.  Our main focus is the small $k$ behavior of $\delta(k)$,
and in particular, the $s$-wave scattering length,
\begin{equation} \label{SCa}
    a \equiv - \lim_{k \to 0} \frac{\tan \delta(k)}{k} . 
\end{equation}
To obtain the correct asymptotic form of $g(r)$, we require that $k^2
< v$.  This is the ``non-leakage'' condition; otherwise the closed
channel is asymptotically open.  We also want to include the two cases
where the closed channel potential $v_c(r)$ can be either attractive
or repulsive, i.e., $\Delta$ can be either positive [attractive
$v_c(r)$] or negative [repulsive $v_c(r)$].  Accordingly we define,
\begin{equation} \label{SC3}
    \kappa(k) \equiv \sqrt{v-k^2} >0, \ \ \alpha(k) \equiv
    \sqrt{k^2-u} \equiv \sqrt{k^2-v+\Delta}, \ (u<k^2) , \beta(k)
    \equiv \sqrt{u-k^2} \equiv \sqrt{v-\Delta-k^2}, \ \ (u>k^2) .
\end{equation} 
The scattering problem requires the solution of the coupled
Schr\"odinger equations,
\begin{eqnarray}
&& -\frac{d^2}{dr^2} f(r)+\tau \delta(r-R) g(r)+[v_c(r)+k^2]f(r)=0, \
\ f(0)=0, \label{SC2a}\\
%\mbox{(Equation for the closed channel wave function)}, \nonumber \\
&& -\frac{d^2}{dr^2} g(r)+\tau \delta(r-R) f(r)+k^2g(r)=0,  \ \
g(0)=0. \ \
%\mbox{(Equation for the open channel wave function)}. 
\label{SC2}
\end{eqnarray}
with the following boundary conditions. 
\begin{equation} \label{SC4}
    f(r) = 
    \begin{cases} A_f e^{-\kappa r} \ \ (r>R) \\ B_f \sin \alpha
    r \ \ (r<R, u<k^2), \\
    B_f \sinh \beta r \ \ (r<R, u>k^2), \end{cases} \ \
    g(r)=\begin{cases} A_g \sin(k r+\delta) \ \ (r>R) \\ B_g \sin k r \
    \ (r<R), 
    \end{cases}
\end{equation}
The four equations for the components of the vector $X \equiv 
(A_f,B_f,B_g,A_g)$ of unknown coefficients are obtained by matching the
wave function and its derivative at $r=R$,
\begin{subequations}
\begin{eqnarray} 
&& f(R^+)=f(R^-) \ \Rightarrow A_f e^{-\kappa R}-B_f \sin \alpha R=0, \label{a1} \\
&& -f'(R^+)+f'(R^-)+\tau g(R)=0 \ \Rightarrow A_f\kappa e^{-\kappa R} +B_f \alpha \cos \alpha R+\tau B_g \sin{kR}=0, \label{a2} \\
&& -g'(R^+)+g'(R^-)+\tau f(R)=0 \ \Rightarrow -A_g k  \cos (kR+\delta) +B_g k \cos k R+\tau B_f \sin \alpha R=0, \label{a3} \\
&& g(R^+)=g(R^-) \ \Rightarrow A_g \sin(kR+\delta)-B_g \sin k R=0. \label{a4} 
\end{eqnarray}
\end{subequations} 
Using trigonometric identities and dividing all coefficients by 
$A_g \cos \delta$ we obtain,
\begin{subequations}
\begin{eqnarray} 
&& A_f e^{-\kappa R}-B_f \sin \alpha R=0, \label{b1} \\
&&  A_f\kappa e^{-\kappa R} +B_f \alpha \cos \alpha R+\tau B_g 
\sin{kR}=0 , \label{b2} \\
&&  - k  \cos kR + k \sin kR \tan \delta +B_g k \cos k R+\tau B_f 
\sin \alpha R=0 , \label{b3}\\
&& \sin kR+\cos kR \tan \delta-B_g \sin k R=0 . \label{b4}
\end{eqnarray}
\end{subequations} 
Denoting $Y \equiv (A_f,B_f,B_g,\tan \delta)$, this can be written as
a set of four linear inhomogeneous equations, ${\cal M}(q) Y=U$:
\begin{equation} \label{SC5}
{\cal M}(q)=\begin{pmatrix}
%------------------First row-------------------------
e^{-\kappa R}&- \sin \alpha R&0& 0 \\
%----------------Second row---------------------
\kappa e^{-\kappa R}&\alpha \cos \alpha R& \tau \sin k R&0 \\
%---------------Third row------------------------
0& \tau \sin \alpha R &k \cos kR & k \sin kR \\
%-----------Fourth row------------------------
0&0&-\sin kR &\cos kR 
\end{pmatrix}, \ \ U = \begin{pmatrix} 0\\0\\k \cos k R \\ -\sin kR \end{pmatrix}. 
\end{equation}
The above set of equations assumes $k^2>u=v-\Delta$, namely
$\alpha=\sqrt{k^2-u}=\sqrt{k^2+\Delta-v} \in \mathbb{R}$.  Otherwise,
$\alpha \to i \beta$ is pure imaginary.  It is easy to check that the solution
for $\tan \delta(k)$ remains real in both cases as long as $k^2<v$,
i.e., the condition that the closed channel is indeed closed.
Simple algebraic manipulations yield the following close expression for 
$\tan \delta(k)$,
\begin{equation} \label{tan_delta}
    \tan \delta(k)=\frac{ \tau^2 \sin kR}{D},
\end{equation}
where we have defined
\begin{equation} \label{SC6}
    D(k,v,\Delta,\tau)= \gamma(k)[F(k)+\kappa(k)]- \tau^2 \cos kR ,
    \quad F(k)=
    \begin{cases}
    \alpha(k) \cot [\alpha(k) R] \ (k^2>u), \\ \beta(k) \coth
    [\beta(k) R] \ (k^2<u),
    \end{cases}
\end{equation}
and $\gamma(k) \equiv k/\sin kR \to 1/R$ as $k \to 0$.

%---------------------------FR at threshold-------------------
\subsection{FR at Threshold}

A {\em FR at threshold} occurs if and only if,
\begin{equation} \label{FRT}
    |\tan \delta(0)|=\infty ,
\end{equation}
This is a slightly modified definition of {\em zero energy resonance}
related to Levinson's theorem in potential scattering \cite{GW,Ballentine,BA}.  
The modification is twofold.  First, we do not require $\delta(\infty)=0$,
as in the formulation of Levinson's theorem, because taking $k^2 \to \infty$
implies that the closed channel leaks current.  Therefore, the scattering
energy is bounded by $k^2<v$ as stated above.  In this region,
$\delta(k)$ is a monotonic function of $k$ and $\delta(0)=(n+1/2)
\pi/2$, where $n$ is an integer (it can also be negative).  The second
modification is that there are situations where $\delta(0)=-\pi/2$ and
$\left [\frac{d \delta(k)}{d k} \right ]_{k=0} > 0$.  Consequently,
$a=+\infty$.  This feature is not encountered in ordinary
potential scattering.
% , but we will not pursue it here.
Following the definition (\ref{SCa}), $a= - {\mathrm{sgn}} \, \delta(0) 
\times \infty$.
  
\begin{figure}[!ht]
\centering
\centering\subfigure[]{\includegraphics[width=0.32\textwidth,angle=0]{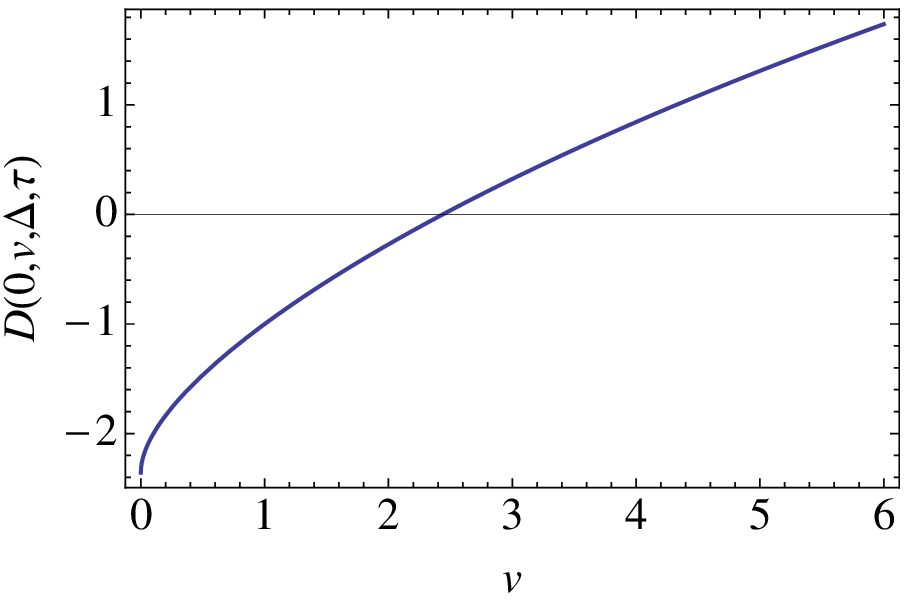}}
\centering\subfigure[]{\includegraphics[width=0.32\textwidth,angle=0]{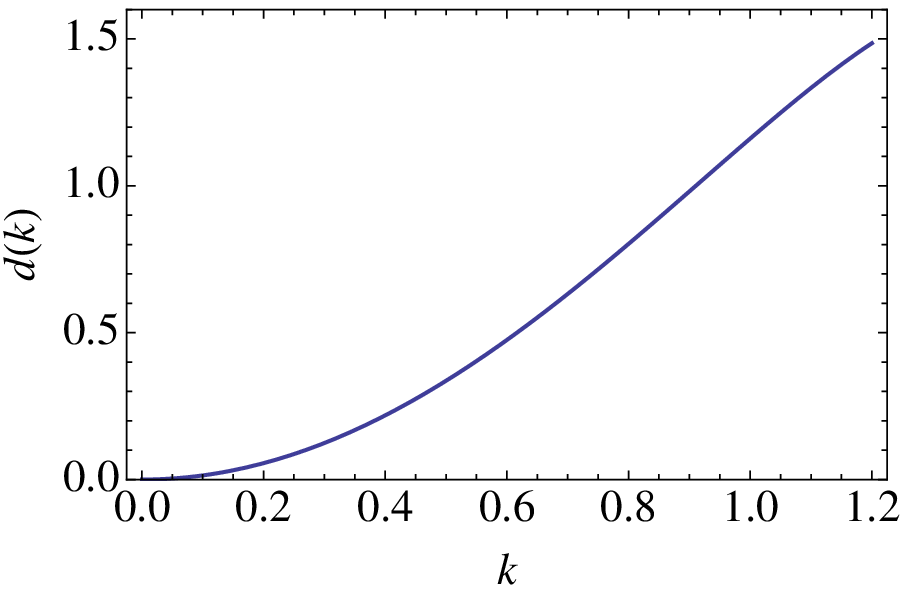}}
\centering\subfigure[]{\includegraphics[width=0.32\textwidth,angle=0]{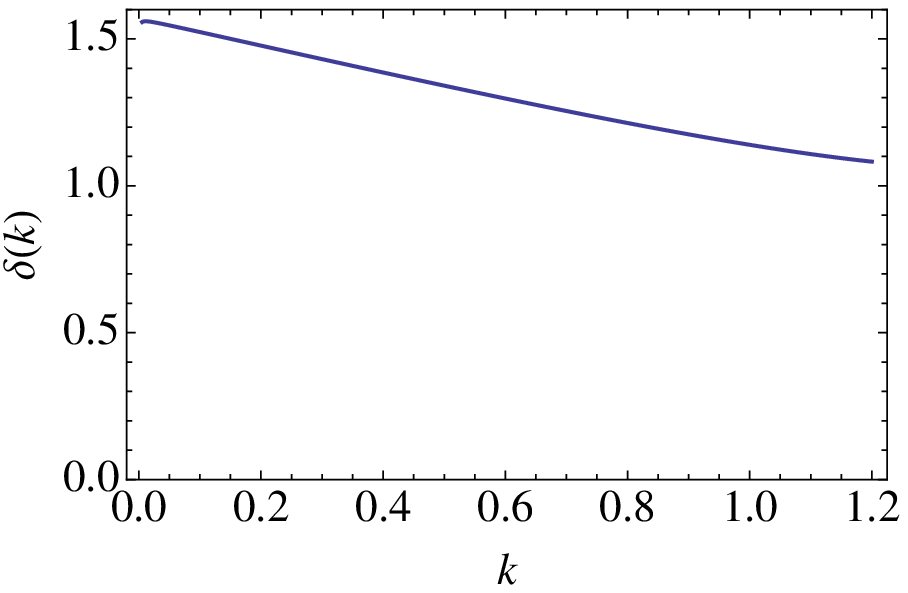}}
\caption{Continuum model with $R=1, \Delta=1, \tau^2=3.0$.  (a)
Graphical solution of equation (\ref{SC7b}) yielding $v_0=2.4378>
\Delta$.  (b) $d(k)$ defined in Eq.~(\ref{dk}), as function of $k$,
showing $d(0)=0$ and $d''(0)>0$ namely $d(k) \approx Ck^2$ for small
$k$ where $C=(1/2) d''(0)>0$.  (c) Phase shift $\delta(k)$ as function
$k$.  The fact that $\delta(0)=\pi/2$, indicates a FR at threshold.
Note that $4 \Delta R^2=4< \pi^2$, namely, the closed channel (when
uncoupled) does not have a bound-state.  And yet, there is a FR in the
coupled systems.}
\label{Fig_Suppl_C123}
\end{figure}

The tunable parameter in this model that drives the system into a FR,
is the potential $v$.  Our first task is then to find the value $v =
v_0(\Delta,\tau)$ for which $D$ vanishes at $k=0$.  Following
Eq.~(\ref{SC6}) we then have to solve the transcendental equations
\begin{subequations}
\begin{eqnarray}
&& \sqrt{\Delta-v} \cot R \sqrt{\Delta-v}+\sqrt{v}-R \tau^2=0 \ (v < \Delta), 
\label{SC7a}\\
&& \sqrt{v-\Delta} \coth R \sqrt{v-\Delta} +\sqrt{v}-R \tau^2=0, \ (v> \Delta) .
\label{SC7b}
\end{eqnarray} 
\end{subequations}
These equations can be solved graphically.  Tuning $v \to v_0(\Delta,
\tau)$ is obviously a necessary condition for FR at threshold.  Now we
show that it is a sufficient condition for $|\tan \delta(0)|=\infty$.
Consider the function
\begin{equation} \label{dk}
 d(k; \Delta, \tau) \equiv D(k,v_0(\Delta, \tau),\Delta,\tau). 
\end{equation} 
$d(k)$ is an even function of $k$ and, by definition $d(0)=0$, and
therefore, for small $k$, $d(k) \approx \tfrac{1}{2}d''(0) k^2$.
Following the expression (\ref{SC6}) for $\tan \delta(k)$ we then have
for small $k$ the following two possibilities:\\
(1) If $v \ne v_0(\tau, \Delta)$ then $D(0,v,\Delta,\tau) \ne 0$ and
$\tan \delta(0)=0, \ \Rightarrow \ \delta(0)=n \pi, \ n=0,
\pm1,\ldots$.  \\
(2) For $v=v_0(\Delta,\tau)$ we have $\tan \delta(k) \approx 2 \tau^2
\sin k R/ (d''(0) k^2) \Rightarrow |\tan \delta(0)|$=$\infty, \
\Rightarrow \ \delta(0)$=$(n+\tfrac{1}{2}) \pi, \ n$=$0,
\pm1,\ldots$.\\
To prove that FR may exist even when the closed channel does not have
a bound-state we recall the elementary analysis of the Schr\"{o}dinger
equation for spherical square well of depth $\Delta$ and range $R$.
The first bound-state enters when $4 \Delta R^2=\pi^2$.  Once we
choose parameters such that $4 \Delta R^2< \pi^2$ and get
$|\delta(0)|=\pi/2$ our task is complete.  This is indeed the case:
The phase-shift $\delta(k)$ obtained from the solution of the
scattering problem with $4 \Delta R^2 < \pi^2$ that has
$\delta(0)=\pi/2$ is shown in Fig.\ref{Fig_Suppl_C123}(c).  This
indicates that there is a FR even when the uncoupled closed channel
does not have a bound-state.

%-----------------------Bound-States in the Continuum---------------------
\subsection{Bound-States of the Coupled System}

\begin{figure}[!ht]
\centering
 \includegraphics[width=0.5 \textwidth,angle=0]{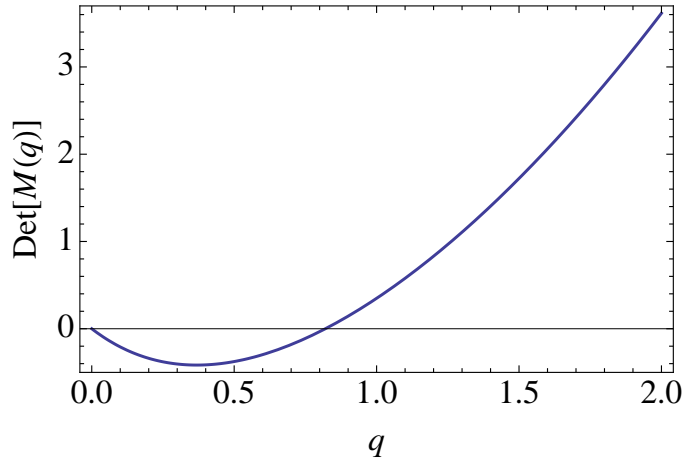}
\caption{Det$[{\cal M} (q)]$ as function of $q$, where ${\cal M} (q)$
is given in Eq.~(\ref{BSC5}).  The parameters used are
$(v,u,\tau,a)=(4,4.2,3,1)$.  Note that $v-u=-0.2$, so that the
spherical well $v_c(r)$ of the closed channel is repulsive (it has
negative ``depth'' $\Delta=v-u$).  And yet, the coupled system has a
bound-state at energy $E_B=-q_B^2$ where $q_B \approx 0.81$.
Performing graphical evaluation of $v_0$ for the relevant parameters
[as in Fig.~\ref{Fig_Suppl_C123}(a)] we find that $v \ll v_0$, i.e.,
the bound-state occurs for $v<v_0$ as it should be.}
\label{Fig_Suppl_BSC}
\end{figure}

Next we consider the bound-state in the continuum model specified by
Eqs.~(\ref{SC2a}) and (\ref{SC2}) and show that there are bound-states
of the coupled system even if the closed channel potential is
repulsive.  To show this, we need to solve the coupled Schr\"odinger 
equations and find a negative energy eigenvalue and square integrable wave
function,
\begin{eqnarray}
&& -\frac{d^2}{dr^2} f(r)+\tau \delta(r- R) g(r)+[v_c(r)+q^2]f(r)=0, \ \ f(0)=0, \ \ \mbox{(Equation for the closed channel wave function)}, \nonumber \\
&& -\frac{d^2}{dr^2} g(r)+\tau \delta(r- R) f(r)+q^2g(r)=0, \ \ g(0)=0, \ \ \mbox{(Equation for the open channel wave function)},
\label{BSC2}
\end{eqnarray}
where $-q^2=\veps_B<0$ is the bound-state energy. Denoting 
\begin{equation} \label{BSC3}
\kappa \equiv \sqrt{v+q^2} >0, \ \ \alpha \equiv \sqrt{u+q^2} \equiv 
\sqrt{v-\Delta+q^2} \ \ \mbox{(either positive or pure imaginary)} ,
\end{equation} 
We have
\begin{equation} \label{BSC4}
f(r)=\begin{cases} A_f e^{-\kappa r} \ \ (x > R) \\ B_f \sinh \alpha r \ \ (r < R), \end{cases}, \ \ 
g(r)=\begin{cases} A_g e^{-q r} \ \ (x > R) \\ B_g \sinh q r \ \ (r < R), \end{cases}
\end{equation}
Now we obtain four homogeneous equations for the components of the
vector $X \equiv(A_f,B_f,B_g,A_g)$ of unknown coefficients by matching
the wave function and its derivative at $r=R$,
\begin{subequations}
\begin{eqnarray} 
&& -f'(R^+)+f'(R^-)+\tau g(R)=0 \ \Rightarrow A_f\kappa e^{-\kappa R}
+B_f \alpha \cosh \alpha R+\tau A_g e^{-q R}=0, \label{1} \\
&& f(R^+)=f(R^-) \ \Rightarrow A_f e^{-\kappa R}-B_f \sinh \alpha R=0,
\label{2} \\
&& g(R^+)=g(R^-) \ \Rightarrow A_g e^{-q R}-B_g \sinh q R=0, \label{3}
\\
&& -g'(R^+)+g'(R^-)+\tau f(R)=0 \ \Rightarrow A_g q e^{-q R} + B_g q
\cosh q R+\tau A_f e^{-\kappa R}=0.  \label{4}
\end{eqnarray}
\end{subequations} 
Writing this set of equations as ${\cal M}(q) X=0$ the matrix ${\cal
M}(q)$ is given by,
\begin{equation} \label{BSC5}
{\cal M}(q)=\begin{pmatrix}
%------------------First row-------------------------
\kappa e^{-\kappa R}&\alpha \cosh \alpha R&0& \tau e^{-q R} \\
%----------------Second row---------------------
 e^{-\kappa R}&-\sinh \alpha R&0&0 \\
%---------------Third row------------------------
0&0&-\sinh q R&e^{-q R} \\
%-----------Fourth row------------------------
\tau e^{-\kappa R}&0&q \cosh q R&q R^{-q R} .
\end{pmatrix}
\end{equation}
A zero of the determinant at $q_B>0$ indicates a bound-state of the
coupled system with bound-state energy $\veps_B=-q_B^2$.  An example
is given in Fig.~\ref{Fig_Suppl_BSC}.
% Thus, we have achieved our goal. 

%---------------------------Proof of positivity-------------------------------
\section{II. The model defined by Eq. (1): Singularity of the $T$ matrix}

% We demonstrated that in the tight binding model and in the continuous
% model based on equation (3) of the main text, FR can occur even when
% the closed channel does not have a bound-state, or even when the
% closed channel potential is repulsive.
% 
We now examine a general two coupled-channel $s$-wave scattering
problem in a 3D continuous geometry in which the open and closed
channel have only a continuous spectrum.  Under these conditions we
prove, that this striking result {\em is not an artifact of the TB
model} or of any other specific model.  For this purpose, it is
convenient to use the language of formal scattering theory without
resorting to a specific representation.  Let ${\cal H}_o$ and ${\cal
H}_c$ denote the Hilbert spaces for the open and closed channel
respectively.  The corresponding Hamiltonians, $H_o$ and $H_c$, act in
${\cal H}_o$ and ${\cal H}_c$, whereas the coupling term $w$ can be
considered as a mapping from ${\cal H}_o$ to ${\cal H}_c$.  For a
given scattering energy $\veps$, the corresponding Green's functions
are denoted as $G_o(\veps) = (\veps+i \eta-H_o)^{-1}$ and
$G_c(\veps)=(\veps+i\eta-H_c)^{-1}$ where $\eta \to 0^+$ guarantees
the proper outgoing asymptotic conditions of the scattering states
in configuration space.
 
Starting from Eq.~(1) of the main text and eliminating the closed
channel component $u_c$ we obtain a single equation for the open channel
component $u_o \in {\cal H}_o$ with an effective (energy dependent)
potential $v_{\mathrm{eff}}$,
\begin{equation} \label{Proof1}
    [H_o+v_{\mathrm{eff}}(\veps)]u_o=\veps u_o , \quad
    v_{\mathrm{eff}}(\veps) = w^\dagger G_c(\veps) w .
\end{equation}
The $T$ matrix associated with $v_{\mathrm{eff}}(\veps)$ is,
\begin{equation} \label{Proof2}
    T(\veps) = v_{\mathrm{eff}}(\veps)[1-G_o(\veps)
    v_{\mathrm{eff}}(\veps)]^{-1}.
\end{equation}

To get the scattering length from the $T$ matrix we denote by $\{ |k
\ra \}$ and $\{ e_k \ge 0 \}$ the set of eigenstates and eigenvalues
belonging to the continuous spectrum of $H_o$ (recall that in general,
$H_o$ may also have a discrete spectrum).  By definition, the $s$-wave
scattering length is given by,
\begin{equation} \label{Proof3}
    a = -C \lim_{e_k \to 0} \la k |T(e_k)|k \ra ,
\end{equation} 
where $C>0$ is a kinematic constant.  Accordingly, an infinite $|a|$,
the hallmark of a FR, implies a singularity of the on-shell $T$ matrix
element as $e_k \to 0$.  Our goal of proving that there is a FR even
when the closed channel does not have a bound-state will be achieved
by verifying the existence of a singularity of the $T$ matrix.

Let us first inspect the case where the closed channel does have a
bound-state $|B \ra$ with binding energy $\veps_B$.  (Recall that
$\veps_B$ is tunable, e.g., by a magnetic field).  The occurrence of
FR is then dominated by this bound-state as discussed in
Ref.~\cite{CCT} where it is assumed that $G_c(\veps)= |B\ra \la B|/(\veps -
\veps_B)$.
% and this leads to his expression (2.24) for the $T$ matrix (what he 
% calls $V_{\mathrm{eff}}$). 
The singularity of $v_{\mathrm{eff}}(\veps)$ at $\veps = \veps_B$ is
slightly shifted upward in the $T$ matrix denominator due to
% the factor $(1$-$G_ov_{\mathrm{eff}})^{-1}$ representing 
virtual transitions to the open channel subspace (self energy).  This
is the reason that, usually, FR occurs at zero scattering energy
$\veps=0^+$ for $\veps_B > 0$ as discussed in the main text.

Now suppose that both $H_c$ and $H_o$ only have a continuous spectrum,
$\{ \epsilon_p \ge v_c(\infty)>0 \}$ (eigenvalues of $H_c$) and, as
already specified, $\{ e_k \ge 0 \}$ (eigenvalues of $H_o$).  Our
strategy is to show that for a given negative energy $\veps$, the
system can be tuned by varying the strength $t$ of the coupling $w$ to
obtain a bound-state at that energy.  In particular, for $\veps=0^-$
(meaning infinitesimally smaller than 0) we can thereby obtain a
zero-energy bound-state.  Then we can employ the fact that by a small
upward shift of $v_c$, a zero energy bound-state at $\veps=0^-$
becomes a zero-energy FR at $\veps=0^+$ (see the Jost function
discussion in the main text).

In Eq.~(\ref{Proof2}) for the $T$ matrix, it is evident that
$v_{\mathrm{eff}}(\veps)$ is regular as $\veps \to 0^\pm$ because
$\veps$ is outside the spectrum of $H_c$ defined by $0 \le \epsilon_p
\le \infty$.  Therefore, if there is a singularity of $\la k |T(e_k)|k
\ra$ leading to an infinite scattering length, or a singularity of
$\la k |T(\veps=0^-))|k \ra$ that corresponds to a zero-energy
bound-state, it must be associated with the second factor of the $T$
matrix,
\begin{equation} \label{Proof4}
    [1-Q(\veps) ]^{-1} \equiv [1 - G_o(\veps)
    v_{\mathrm{eff}}(\veps)]^{-1} = [1-G_o(\veps) w^\dagger
   G_c(\veps)w] .
\end{equation}
Thus, we need to show that $[1-Q(\veps = 0^-) ]^{-1} $ can be tuned to
singularity.  For this purpose, it is sufficient to show that $Q(\veps
= 0^-)$ acting in ${\cal H}_o$ is positive definite, because then, the
operator $[1-Q(\veps =0^-)]^{-1}$ can be tuned to be singular (by
varying the strength $t$ of the coupling $w$), leading to a pole of
$\la k |T(\veps))|k \ra$ at $\veps=0^-$.

Sine both $H_o$ and $H_c$ have a positive continuous spectrum, the
negative energies $\veps<0$ are outside the spectra of both $H_o$ and
$H_c$.  Hence, there is no need to add a small imaginary part $i \eta$
to the energy variable because the limit $\eta \to 0$ can be safely
applied.  In other words, for $\veps<0$,
\begin{equation} \label{Proof5}
    Q(\veps)=(\veps-H_o)^{-1} w^\dagger (\veps-H_c)^{-1}w .
\end{equation}
Consider the eigenvalue equation defined for $| \psi \ra \in {\cal
H}_o$,
\begin{equation} \label{Proof5}
    Q(\veps)|\psi \ra=(H_o-\veps)^{-1}w^\dagger (H_c-\veps)^{-1} w
    |\psi \ra = \lambda |\psi \ra~.
\end{equation}
(Note that the double sign changes in the resolvents compensate each
other).  We want to show that as $\veps \to 0^-$, $\lambda>0$.
Clearly, for a given $\veps < 0$, both denominators are positive
definite.  Therefore, the operators $(H_o-\veps)^{-1/2}$ and
$(H_c-\veps)^{-1/2}$ are both Hermitian.  Using the notation, $|\psi
\ra = (H_o-\veps)^{-1/2}|\phi \ra$ the eigenvalue problem becomes
Hermitian, i.e.,
\begin{equation} \label{Proof6}
    \underbrace{(H_o-\veps)^{-1/2} w^\dagger
    (H_c-\veps)^{-1/2}}_{A^\dagger} \underbrace{(H_c-\veps)^{-1/2} w
    (H_o-\veps)^{-1/2}}_A |\phi \ra=\lambda |\phi \ra~.
\end{equation}
Moreover, the Hermitian operator acting on $|\phi \ra$ has the form
$A^\dagger A$ where $A=(H_c-\veps)^{-1/2} w (H_o-\veps)^{-1/2}$,
therefore, it is positive definite and hence $\lambda >0$.  Thus, we
have established that the system can be tuned to have a zero-energy
bound-state.  By an infinitesimally small upward shift of $v_c$ this
bound-state becomes a pole of the on-shell $T$ matrix $\la k|T(e_k)|k
\ra$ as $k \to 0$, namely, a FR.
%%%%%%%%%% sign of a at unitarity %%%%%%%%%%
\section{III. Relation between coupling strength $\tau$ and the sign
of $a$ near unitarity}

From Eq.~(2) we see that Sign$[a]=-$Sign$[\delta(0)]$.  To elucidate
the sign of $\delta(0)$ let us return to Eq.~(11) of the main text,
$\tan \delta = N(k,\tau,v_0,\Delta,\Lambda) / D(k, \tau, v_0, \Delta,
\Lambda)$ where $v_0$ is defined in Eq.~(12).  Since $N(k...)=C \sin k
\ \ (C>0)$, the sign of $a$ is determined by the sign of $D(k...)$ as
$k \to 0$.  It is more convenient to express $D$ as function of
$\veps_k$ and inspect its sign as $\veps_k \to -2$.  Explicitly,
\begin{eqnarray} \label{sign1}
&& D(\veps_k,\tau,v_0,\Delta,\Lambda)=\sqrt{(v_0-\veps_k)^2-4} + 
v_0-2 \Delta+(\tau^2-1)\veps_k \nonumber \\
&& \quad + \tfrac{1}{2} \Lambda \veps_k [(v_0-2 \Delta)   +
\sqrt{(v_0-\veps_k)^2-4}  -\veps_k]~,
\end{eqnarray}
where $v_0$, defined in Eq.~(12) of the text is such that
$D(-2,\tau,v_0,\Delta,\Lambda)=0$.  Expanding $D(\veps_k, \tau, v_0,
\Delta, \Lambda)$ near threshold $\veps_k=-2$ and recalling that, by
construction, $D(-2,\tau,v_0,\Delta,\Lambda)=0$ we get,
\begin{equation} \label{sign2}
    D(\veps_k,\tau,v_0,\Delta,\Lambda) \approx
    D'(-2,\tau,v_0,\Delta,\Lambda)(\veps_k+2) = 4D'(-2,\tau, v_0,
    \Delta, \Lambda)\sin^2\tfrac{k}{2}~.
\end{equation}
This result shows that (1) the denominator of $\tan \delta(k)$
vanishes faster that its numerator $N(k...)=C \sin k$, as discussed
after equation (12) of the text.  (2) The sign of $\delta(0)$ is
determined by the sign of $D'(-2,\tau,v_0,\Delta,\Lambda)$.  For fixed
$\Lambda$ and $\Delta$, the function $f(\tau) \equiv
D'(-2,\tau,v_0,\Delta,\Lambda$ is a complicated function of $\tau$
(recall that $v_0$ as given by Eq.~(12) of the text also depends on
$\tau$).  However, it is not difficult to check that $f(\tau)$ changes
sign at some point $\tau_0$ such that ${\mathrm{Sign}}[f(\tau)] =
{\mathrm{Sign}}[\delta(0)] = {\mathrm{Sign}}[(\tau-\tau_0)]$.  Thus we
have,
\begin{equation} \label{sign3}
    {\mathrm{Sign}}[a] = - {\mathrm{Sign}}[\delta(0)] =\begin{cases} +
    \ \ (\tau<\tau_0) \\ - \ \ (\tau > \tau_0) \end{cases}.
\end{equation} 
In actuality, $\tau_0$ is only weakly dependent on $\Delta$ and
$\Lambda$.

 %\end{multicols}
\end{document}